\documentclass[aps,prd,10pt,notitlepage,nofootinbib,superscriptaddress,showkeys,showpacs]{revtex4-2}
\pdfoutput=1
\usepackage{amstext,amsmath,amssymb,amsfonts,bbm,euscript,color,dsfont}
\usepackage[latin1]{inputenc}
\usepackage{epsfig}
\usepackage{hyperref}
\usepackage{amsthm}
\usepackage{color}
\usepackage{slashed}
\usepackage{dashrule}
\usepackage{xcolor}
\usepackage{physics}
\usepackage{ulem}

\usepackage{latexsym}

\usepackage{cancel}

\newcommand{\bea}{\begin{eqnarray}}	
\newcommand{\eea}{\end{eqnarray}}
\newcommand{\be}{\begin{equation}}	
\newcommand{\ee}{\end{equation}}
\newcommand{\beq}{\begin{equation}}	
\newcommand{\eeq}{\end{equation}}

\newcommand{\Z}{{\mathbb Z}}
\newcommand{\C}{{\mathbb C}}

\def\R{\relax\ifmmode {\mathbb R}  \else${\mathbb R}$\fi}
\def\C{\relax\ifmmode {\mathbb C}  \else${\mathbb C}$\fi}
\def\Z{\relax\ifmmode {\mathbb Z}  \else${\mathbb Z}$\fi}
\def\N{\relax\ifmmode {\mathbb N}  \else${\mathbb N}$\fi}
\def\I{\relax\ifmmode {\mathbb I}  \else${\mathbb I}$\fi}

\begin{document}
\title{The Euler-Heisenberg action for a $U(1)\times U(1)$ dyon quantum electrodynamics} 

\author{D.O.R. Azevedo}\email{azevedo.dor@gmail.com}
 \affiliation{Instituto de F\'isica, Facultad de Ingenier\'ia, Universidad de la Rep\'ublica, J. H. y Reissig 565, 11000 Montevideo, Uruguay}
\author{T.S. Dias} \email{thadeu.dias@ufv.br}\thanks{(Corresponding author)}
\affiliation{Departamento de F\'isica, Universidade Federal de Vi\c cosa,
	Campus Universit\'ario, Avenida Peter Henry Rolfs s/n, 36570-900, Vi\c cosa, MG, Brazil}
 \affiliation{Ibitipoca Institute of Physics (IbitiPhys), Concei\c c\~ao do Ibitipoca, 36140-000, MG, Brazil}
\author{E.D. Pereira} \email{emilio.drumond@ufv.br}
\affiliation{Departamento de F\'isica, Universidade Federal de Vi\c cosa,
	Campus Universit\'ario, Avenida Peter Henry Rolfs s/n, 36570-900, Vi\c cosa, MG, Brazil}
 \affiliation{Ibitipoca Institute of Physics (IbitiPhys), Concei\c c\~ao do Ibitipoca, 36140-000, MG, Brazil}

\begin{abstract}
The one-loop effective Lagrangian of quantum electrodynamics
for dyons (dQED) with a $U(1) \times U(1)$ gauge symmetry is
derived using the Schwinger proper-time method. We identify the
analog of the Schwinger pair-production limit and compute the corresponding nonlinear equations of motion. The electromagnetic response of the model is analyzed by linearizing the equations of motion around a purely magnetic background. From the resulting plane-wave solutions, we obtain the effective permittivity and permeability tensors, along with the associated dispersion relations and refractive indices for parallel and perpendicular polarization modes. The refractive indices involve hybrid superpositions of both gauge sectors, as illustrated in the symmetric case $q_{(1)} = q_{(2)}$. In this background, the model exhibits vacuum birefringence, indicating that the quantum vacuum behaves as an anisotropic medium. All results consistently reduce to the standard Euler-Heisenberg predictions of QED when the magnetic charge vanishes.
\end{abstract}

\maketitle

\textit{In honor of the 60th birthday of Prof. Oswaldo Monteiro Del Cima}
\section{Introduction}
Electromagnetism stands as one of the most remarkable achievements in the history of physics. Through the seminal work of Maxwell, electricity and magnetism were unified into a single theoretical framework, thereby revealing the intrinsic connection between the two. Since then, Maxwell's equations have constituted a cornerstone of both classical and quantum physics, governing a wide range of physical phenomena. Nevertheless, their asymmetry with respect to electric and magnetic sources has long been a source of conceptual dissatisfaction. Why would nature favor electric charges over their magnetic counterparts? Why do magnetic monopoles not manifest as fundamental sources analogous to the electric charge? Such questions have motivated extensive theoretical and experimental efforts and have led to one of the most enduring open problems in modern physics: the existence of the magnetic monopole. 

In 1931, Dirac showed that the existence of magnetic monopoles was compatible with quantum mechanics and that it implied a fundamental relation between electric and magnetic charges \cite{Dirac:1931kp}. This relation required that the product of the charges must be proportional to an integer number ($qg=n/2,\, n\in\mathbb{Z}$), explaining the observed quantization of electric charges as multiples of the electron charge $e$. However, it introduced an unobservable singularity in the vector potential connecting monopoles of opposite charges, which came to be known as the Dirac string. The theory was later reformulated in more general settings \cite{Dirac:1948um}, while maintaining the previous features.
From the 1960s onward, the topic received contributions from several distinct approaches, such as the works of Schwinger~\cite{Schwinger:1966nj,Schwinger:1967rg,Schwinger:1968rq,Schwinger:1969ib} incorporating monopoles into quantum field theory, where the concept of dyons, particles carrying both electrical and magnetic charges, was introduced; to the ideas of monopoles as extended topological configurations in non-abelian gauge theories \cite{tHooft:1974kcl,Polyakov:1974ek,Julia:1975ff}, which were shown to be stable above a certain mass, known as the BPS bound \cite{Bogomolny:1975de,Prasad:1975kr}; the connections with the geometrical formulation of gauge theories in terms of principal fiber bundles \cite{Wu:1975es,Wu:1976ge}; as well as hybrid formulations between Dirac's and topological solutions, like the Cho-Maison monopole \cite{Cho:1996qd,Cho:2013vba}. For a comprehensive review, we refer to \cite{Shnir:2005vvi,Mavromatos:2020gwk}.

One way to avoid the presence of Dirac strings, without deviating much from Dirac's ideas, was to introduce a dual potential associated to the magnetic current \cite{Schwinger:1966nj,Zwanziger:1968rs}. It involved a non-local formulation of the theory, which was later recast in a local way by Zwanziger \cite{Zwanziger:1970hk}. Those proposals usually also involved restrictions similar to Dirac's quantization condition. Other proposals involving two potentials are those of Cabibbo and Ferrari \cite{Cabibbo:1962td} and Salam \cite{Salam:1966bd}. These present remarkable similarities, by modifying the electromagnetic field strength to include both potentials, although Salam's does not impose any kind of quantization condition.
Building upon these ideas, we study a framework describing fermionic dyons in which both electric and magnetic charges are treated independently and perturbatively~\cite{Govaerts:2023iqf}. The corresponding electromagnetic interaction is formulated as a gauge theory with symmetry group $U(1)\times U(1)$, hereafter referred to as a dyon quantum electrodynamics (dQED). This model has been analyzed in detail in Ref.~\cite{Dias:2025xwu}, using the algebraic renormalization formalism~\cite{Piguet:1995er}, where we have demonstrated that the theory is free from gauge and infrared anomalies to all orders in perturbation theory and multiplicatively renormalizable. These results motivate the investigation of dyonic and electromagnetic phenomena within this framework.

Even though they appear in many different theoretical contexts, magnetic monopoles remain as an elusive particle from the experimental and phenomenological points of view. Nevertheless, there is a continued effort to search for them, covering a wide range of approaches, such as monopoles trapped/bound in ordinary matter~\cite{Kolm:1971xb,Alvarez:1970zu,Bendtz:2013tj,Jeon:1995rf,Price:1986ky,Burdin:2014xma}, in cosmic ray detections~\cite{MACRO:2002jdv,IceCube:2021eye,ANTARES:2022zbr}, astrophysical and cosmological observations~\cite{Adams:1993fj,Kobayashi:2023ryr,Zhang:2024mze,Perri:2025qpg,Perri:2025vtn} as well as produced in colliders~\cite{Kalbfleisch:2003yt,H1:2004zsc,Fairbairn:2006gg,OPAL:2007eyf,ATLAS:2023esy,MoEDAL:2023ost}. Moreover, they could appear due to the Schwinger mechanism in strong magnetic fields, analogously to the electron-positron pair production in electric fields~\cite{Affleck:1981ag,Gould:2019myj,Gould:2024zed}. For a review of these searches, see~\cite{Patrizii:2015uea,Rajantie:2024wuw}. 

Due to the difficulties of detecting monopoles,  it is useful to study their effects on the other fields of the theory, which would imply their presence. In our case, it amounts to the two $U(1)$ gauge fields describing the electromagnetic field. The development of theoretical frameworks that reduce (or even eliminate) explicit dependence on matter fields, which have become powerful tools in the search for new physics, is therefore well suited to our purposes. One particularly promising direction is nonlinear electrodynamics (NLED), which extends Maxwell's theory by incorporating self-interactions of the electromagnetic field, of which one of the earliest examples was the work of Euler and Heisenberg~\cite{Heisenberg:1936nmg}. A comprehensive historical discussion of their results and influence on modern developments is provided in ~\cite{Dunne:2012vv}. The Euler-Heisenberg NLED predicts several hallmark nonlinear QED effects, including vacuum polarization with pair production, light-by-light scattering, and vacuum birefringence, indicating that the quantum vacuum behaves as an optically anisotropic medium under a strong magnetic background. Another notable example of a NLED is the Born-Infeld theory~\cite{Born:1934gh}, originally proposed to regularize the divergence of the electron self-energy and now of great interest in string theory~\cite{Fradkin:1985qd,Cederwall:1996uu,Gibbons:1997xz}. Extensions of NLED have been employed in diverse contexts, such as in models describing electrically charged black holes~\cite{Kruglov:2017fuj,Kruglov:2017xmb,kruglov2024heat,Nozari:2025zlt,Channuie:2025xlw,Lambiase:2024lvo}, studies of the inflationary epoch~\cite{Kruglov:2020axn,Benaoum:2022uta}, gamma-ray burst phenomenology~\cite{Brevik:2025fev}, possible connections with relativistic superconductivity~\cite{bruce2024nonlinear,bruce2025nonlinear}, and vacuum polarization in spatially inhomogeneous backgrounds~\cite{Marmier:2024rwd}, among several other applications. 

An approach to nonlinear electrodynamics that takes monopoles into account can be found in the work of~\cite{Kovalevich:1997de}, which provides important insights into the computation of the effective Lagrangian for QED coupled to dyons. In their approach, an effective Lagrangian is obtained within a dyonic framework, leading to nonlinear $P$ and 
$T$-violating terms. These terms contribute to light-by-light scattering and induce a contribution to the electron dipole moment, from which one is able to estimate a lower bound to the dyon mass.

In the present work, we investigate the nonlinear formulation of the $U(1)\times U(1)$ dQED as a one-loop Euler-Heisenberg effective theory. We make use of Schwinger's proper-time formalism~\cite{Schwinger:1951nm}, extended to the case of a theory containing two $U(1)$ gauge fields, which proves to be highly advantageous - not only as a powerful regularization technique, but also as a framework that allows a detailed investigation of gauge invariance and provides a clear formulation of the underlying technical aspects~\cite{Dittrich:2000zu}. From the effective action, we derive a few results, like the Schwinger limit for pair production and modified Maxwell equations. From the latter, we obtain the effective permittivity and permeability tensors, leading to a prediction of magnetic vacuum birefringence. Moreover, all results obtained are consistent with the ones from QED, recovered at the limit of vanishing magnetic charge. It is organized as follows: in section \ref{II}, we present a brief overview of the main properties of the dyon electromagnetism; in section \ref{III}, we compute the effective Lagrangian, the Schwinger limit, and the field equations of the nonlinear dQED. In section \ref{IV}, we
obtain the electromagnetic displacement tensor, which determines the permeability and permittivity tensors, and analyze the dispersion relations for the propagation of the gauge fields in a constant background field. In section \ref{V}, we compute the vacuum birefringence of the theory in a constant background field. Finally, some concluding remarks are presented in section \ref{VI}. We collect the explicit expression for the equations of motion in the appendix \ref{apx:B}.
\section{Dyon electrodynamics}
\label{II}
We define the classical action for a dyon electrodynamics (dQED) with two gauge fields \cite{Cabibbo:1962td,Salam:1966bd,Govaerts:2023iqf,Dias:2025xwu} as
\begin{equation}\label{protoaction}
\Sigma_{dQED}=\int d^{4}x \left[-\frac{1}{4}\mathcal{F}_{\mu\nu}\mathcal{F}^{\mu\nu}+\bar{\psi}(\imath \gamma^{\mu}D_{\mu}-m )\psi \right],
\end{equation}
with the modified electromagnetic field-strength\footnote{We use the convention $\varepsilon^{0123}=-\varepsilon_{0123}=1$.}
\begin{equation}\label{eq10}
\mathcal{F}^{\mu\nu}=\partial^{\mu}A^{\nu}-\partial^{\nu}A^{\mu}+\varepsilon^{\mu\nu\rho\sigma}\partial_{\rho}B_{\sigma},
\end{equation}
and the covariant derivative defined as
\begin{equation}
    D_{\mu}=\partial_{\mu}-\imath eA_{\mu}-\imath gB_{\mu},
\end{equation}
where $e$ and $g$ are the electric and magnetic coupling constants, respectively, and $m$ is the mass of the fermionic field. The dQED has a $U(1)\times U(1)$ gauge symmetry, with each gauge field associated to a different sector of the theory: $A_{\mu}$ being related to the electric charge $e$ and $B_{\mu}$ related to the magnetic charge $g$. The presence of a second potential allows one to avoid the singularities of the Dirac strings and removes the necessity for a quantization condition relating electric and magnetic charges \cite{Salam:1966bd,Govaerts:2023iqf}. However, this new configuration for the electromagnetic field introduces extra degrees of freedom due to the second gauge sector, which we identify as photons and metaphotons. 

The action \eqref{protoaction} can be rewritten as
\begin{equation}\label{eq15}
    \Sigma_{dQED}=\int d^{4}x \left[-\frac{1}{4}F_{\mu\nu}F^{\mu\nu}-\frac{1}{4}G_{\mu\nu}G^{\mu\nu}+\bar{\psi}(\imath \gamma^{\mu}D_{\mu}-m )\psi \right],
\end{equation}
with,
\begin{equation}
   F^{\mu\nu}=\partial^{\mu}A^{\nu}-\partial^{\nu}A^{\mu},~~~~~G^{\mu\nu}=\partial^{\mu}B^{\nu}-\partial^{\nu}B^{\mu}.
\end{equation}
In this form, the two kinds of gauge bosons present in the theory become transparent. 

In a theory including magnetic charges, the usual CPT theorem must be modified to take its effects into account. Namely, the charge conjugation operator now must include the electric as well as the magnetic charge conjugations \cite{Ramsey:1958gvj}. Notice that, to comply with the definition of the total field strength \eqref{eq10}, the quantities associated with the magnetic sector must be pseudo-vectors and pseudo-scalars under parity \cite{Singleton:1996hgp}. Moreover, the action exhibits an additional symmetry that exchanges the electric and magnetic sectors, characterizing a duality symmetry, which is more easily seen when expressed with the two separate field strengths. In fact, the action is more generally invariant under a $SO(2)$ rotation mixing the electric and magnetic sectors, which reduces to a duality transformation exchanging the fields for specific values of the rotation angle $\theta$ \cite{Singleton:1996hgp,Govaerts:2023iqf}.

At tree level, the action was examined through the study of the propagators, showing that the vector fields $A_{\mu}$ and $B_{\mu}$ propagate massless physical degrees of freedom, corresponding to two transverse modes for each photon and metaphoton. As a consequence, the model is free of tachyonic excitations and satisfies the necessary condition for $S$-matrix unitarity, namely the absence of negative-norm states and nonphysical poles in the propagators. Although established at tree level, this result is later confirmed at the quantum level by means of algebraic renormalization \cite{Dias:2025xwu}.

To simplify the notation, we introduce an index $(a)$ to identify each gauge sector, so that:
\begin{align}
&F_{\mu\nu}^{\scalebox{0.57}{(1)}}=F_{\mu\nu},~~~~A_{\mu}^{\scalebox{0.57}{(1)}}=A_{\mu},~~~~~q_{\scalebox{0.57}{(1)}}=-e,\nonumber\\
&F_{\mu\nu}^{\scalebox{0.57}{(2)}}=G_{\mu\nu},~~~~A_{\mu}^{\scalebox{0.57}{(2)}}=B_{\mu},~~~~q_{\scalebox{0.57}{(2)}}=-g,\\
&D_{\mu}=\partial_{\mu}-\imath eA_{\mu}-\imath gB_{\mu}=\partial_{\mu}+\imath q_{(a)}A_{\mu}^{(a)}.\nonumber
\end{align}

From the modified field strength $\mathcal{F}^{\mu\nu}$ in equation \eqref{eq10}, we can identify the original electric and magnetic fields $(\vec{E},\vec{B})$ as~\cite{Govaerts:2023iqf}
\begin{align} \label{fieldsE_B}
&\vec{E}=\vec{E}_{(1)}-\vec{B}_{(2)},~~~~~~~~~~\vec{B}=\vec{B}_{(1)}+\vec{E}_{(2)},
\end{align}
in which each sector's ``electric" and ``magnetic" fields are defined as in Maxwell's electrodynamics, that is, $\vec{E}_{(a)}=-\nabla A^{(a)}_0+\partial_t\vec{A}^{(a)}$ and $\vec{B}_{(a)}=\nabla\times\vec{A}^{(a)}$, respectively.This way, the characteristic feature of dQED becomes explicit: the magnetic sector components contribute directly to the original electromagnetic fields. This is the reason no strings are needed in such a formulation, since the total magnetic field $\vec{B}$ of an static monopole/dyon comes entirely from the magnetic sector field $\vec{E}_{(2)}$, as can be seen in \eqref{fieldsE_B}, and no singular line is needed in the spatial components of $A_\mu^{(1)}$. 
\section{Effective Lagrangian}
\label{III}
The quantum consistency of the dQED was studied in \cite{Dias:2025xwu} to all orders in perturbation theory within the framework of algebraic renormalization~\cite{Piguet:1995er}. It was shown that it is free of gauge anomalies and multiplicatively renormalizable. Since the renormalizability properties of the dQED are well established, we now explore some of the phenomenological consequences of the model, such as the properties of the quantum vacuum of the theory.

To this end, we compute the one-loop effective action for the gauge fields~\cite{Schwartz:2014sze,Dittrich:2000zu}. The resulting theory encodes nonlinear interactions between the gauge fields, which in our case can present as mixing terms between photon and metaphoton fields, and allows us to investigate physical phenomena like pair productions~\cite{Schwinger:1951nm}, vacuum birefringence, light-by-light scattering, etc., separating low energy processes in relation to the energy scale set by the dyon mass $m$. In this context, the derivation of an effective action is particularly well motivated, since the absence of gauge anomalies allows the fermionic fields to be integrated out without obstructions~\cite{Petrov:2016azi,Deser:1997gp,Jacobson:2018kso,Bilal:2008qx,Weinberg:1996kr}. Thus, we have
\begin{align}
e^{\imath\Gamma_{eff}{[A,B]}}&=\int\mathcal{D}\bar{\psi} ~\mathcal{D}\psi~e^{\imath\Sigma_{dQED}}.
\end{align}
The path integral on the right-hand side reduces to a functional determinant of the Dirac operator, such that
\begin{align}
 \Gamma_{eff}[A,B]&=-\frac{1}{4}\int d^{4}x F_{\mu\nu}^{(a)}F_{(a)}^{\mu\nu} - \imath \ln\det (\imath\slashed{D}-m) - \imath \ln\mathcal{N}
\end{align}
where $\Gamma_{eff}$ denotes the effective action and $\mathcal{N}$ is a normalization constant arising from the fermionic path integral.

In order to obtain a well-defined effective action, the functional determinant of the Dirac operator must be properly defined and evaluated using an appropriate regularization scheme. The computation of functional determinants amounts to the evaluation of all one-loop 1PI Feynman diagrams with one closed fermion loop, and their regularization corresponds to the regularization of the fermionic propagator, which must be implemented in a gauge-invariant manner. Once a regularization scheme for the fermion propagator is chosen, it consistently induces the same regularization for all fermionic determinants~\cite{Bilal:2008qx,Weinberg:1996kr}.

In the present work, we use a direct approach to calculating the effective action $\Gamma_{eff}$, based on Schwinger's proper time method \cite{Schwinger:1951nm} applied to the Feynman path integral \cite{Schwartz:2014sze}. We integrate out the fermionic fields in the functional generator assuming constant background electromagnetic fields. Applying the identity $\ln \det A = \Tr \ln A$ and taking the derivative of the term coming from the determinant with respect to $m^2$, we can use Schwinger's parametrization\footnote{We omit the $i\varepsilon$ dependence in the final expression, which comes from the Feynman prescription to the fermion propagator and is crucial in Schwinger's parametrization, throughout the text.} and then integrate back over $m^2$, yielding
\begin{align}
    \ln\det (\imath\slashed{D}-m) = -\frac{1}{2}\int d^4x \int_{0}^{\infty}\frac{ds}{s}e^{-\imath sm^{2}}\Tr\left[\bra{x}e^{-\imath s\slashed{D}^{2}}\ket{x}\right] + cte
\end{align}
where $\Tr$ indicates a Dirac trace and $cte$ is an integration constant. Therefore, the effective Lagrangian can be written as
\begin{equation}
    \mathcal{L}_{eff}=-\frac{1}{4}F_{\mu\nu}^{(a)}F_{(a)}^{\mu\nu}+\frac{\imath}{2}\int_{0}^{\infty}\frac{ds}{s}e^{-\imath sm^{2}}\Tr\left[\bra{x}e^{-\imath s\slashed{D}^{2}}\ket{x}\right]+cte.
\end{equation}
where we combined in $cte$ the normalization and integration constants, which do not affect the physical content of the theory and will be useful in the definition of the properly renormalized effective action.

Following Schwinger's original method, with suitable generalizations to account for the two gauge sectors coupled to the fermion field, we evaluate the proper-time propagation and the trace of the covariant derivative operator. This generalization is straightforward, so we do not present a step-by-step demonstration here\footnote{See Refs. \cite{Schwinger:1951nm, Schwartz:2014sze,Dittrich:2000zu} for further details.}. The presence of a second gauge field yields mixing terms between the sectors, which vanish in the limit $g\to 0$, recovering the results of QED. This will serve as a consistency check throughout all calculations presented here. After evaluating the proper-time propagation, we are left with
\begin{align}\label{leff}
&\mathcal{L}_{eff}=-\frac{1}{4}F_{\mu\nu}^{(a)}F_{(a)}^{\mu\nu}+\frac{1}{32\pi^{2}}\int_{0}^{\infty}\frac{ds}{s^{3}}\Tr \left[\exp\left( -\imath s m^2- \frac{1}{2}tr\ln\left[\frac{\sinh(sq_{(a)}\mathbf{F}^{(a)})}{sq_{(a)}\mathbf{F}^{(a)}}\right]+\frac{\imath s}{2}tr(\boldsymbol{\sigma} q_{(a)}\mathbf{F}^{(a)})\right)\right],
\end{align}
where boldface symbol $\mathbf{F}=F_{\mu \nu}$ denotes the matrix form of a given tensor and $tr$ indicates contraction of space-time indices.
To evaluate the Dirac trace and get the final expression for the Lagrangian, we need the eigenvalues of $\boldsymbol{\sigma} q_{(a)}\mathbf{F}^{(a)}$ and $\mathbf{F}^{(a)}$\footnote{See \cite{Schwinger:1951nm,Schwartz:2014sze} for a complete derivation in the case of QED. Generalization to the case considered here is straightforward due to the linearity of the quantity $q_{(a)}\mathbf{F}^{(a)}$ in each gauge sector and of the trace.}. Each of these eigenvalues are written in terms of two Lorentz scalars for each sector:
\begin{equation}
\mathcal{F}^{(a)}=\frac{1}{4}F_{\mu\nu}^{(a)}F^{\mu\nu}_{(a)}=\frac{1}{2}(B_{(a)}^{2}-E_{(a)}^{2})~~~~ \mathrm{and}~~~~\mathcal{G}^{(a)}=-\frac{1}{4}F_{\mu\nu}^{(a)}\tilde{F}^{\mu\nu}_{(a)}=-\vec{B}_{(a)}\cdot\vec{E}_{(a)}, 
\end{equation}
with ${B}_{(a)}=|\vec{B}_{(a)}|$ and ${E}_{(a)}=|\vec{E}_{(a)}|$. The eigenvalues $\lambda^{tr}$ of $\Tr(\boldsymbol{\sigma}\mathbf{F}^{(a)})$ are
\begin{align}
\lambda^{tr}=\left\{+2\,z^{(a)},-2\,z^{(a)},+2\,\bar{z}^{(a)},-2\,\bar{z}^{(a)}\right\},
\end{align}
with
\begin{align}
z^{(a)}=\sqrt{2\left(\mathcal{F}^{(a)}+\imath\mathcal{G}^{(a)}\right)},~~~~~\bar{z}^{(a)}=\sqrt{2\left(\mathcal{F}^{(a)}-\imath\mathcal{G}^{(a)}\right)}.
\end{align}
After some algebraic manipulation, we get
\begin{align}
\Tr\Big[\exp\left(\frac{\imath s}{2}tr(\boldsymbol{\sigma}q_{(a)}\mathbf{F}^{(a)})\right)\Big]&=2\cos\Big(sq_{(a)}z^{(a)}\Big)+2\cos\Big(sq_{(a)}\bar{z}^{(a)}\Big)\\
&=4\cos\Big(\frac{sq_{(a)}z^{(a)}+sq_{(a)}\bar{z}^{(a)}}{2}\Big)\cos\Big(\frac{sq_{(a)}z^{(a)}-sq_{(a)}\bar{z}^{(a)}}{2}\Big).\nonumber
\end{align}

The eigenvalues $\lambda^{F}$ of $\mathbf{F}^{(a)}$ are given by
\begin{align}
\lambda^{F}=\Big\{+2\lambda^{(a)}_{+},-2\lambda^{(a)}_{+},+2\lambda^{(a)}_{-},-2\lambda^{(a)}_{-}\Big\},
\end{align}
with
\begin{align}
\lambda^{(a)}_{\pm}=\frac{\imath}{2}\Big[(z^{(a)})\pm(\bar{z}^{(a)})\Big].
\end{align}
Summing the second term in the exponential in \eqref{leff} over the eigenvalues $\lambda^F$, we get
\begin{align}
    \frac{1}{2}tr\ln\left[\frac{\sinh(sq_{(a)}\mathbf{F}^{(a)})}{sq_{(a)}\mathbf{F}^{(a)}}\right]=\ln \left[\frac{\sinh(sq_{(a)}\lambda^{(a)}_+)\sinh(sq_{(a)}\lambda^{(a)}_-)}{s^2(q_{(a)}\lambda^{(a)}_+)(q_{(a)}\lambda^{(a)}_-)} \right],
\end{align}
and therefore
\begin{align}
\exp\left(- \frac{1}{2}tr\ln\left[\frac{\sinh(sq_{(a)}\mathbf{F}^{(a)})}{sq_{(a)}\mathbf{F}^{(a)}}\right]\right)=\frac{1}{4}\frac{s^{2}\left[(q_{(a)}z^{(a)})^{2}-(q_{(a)}\bar{z}^{(a)})^{2}\right]}{\Big[\sin\Big(\frac{sq_{(a)}z^{(a)}+sq_{(a)}\bar{z}^{(a)}}{2}\Big)\sin\Big(\frac{sq_{(a)}z^{(a)}-sq_{(a)}\bar{z}^{(a)}}{2}\Big)\Big]}.
\end{align}

With these results, the fermion loop contribution to the effective Lagrangian can be written as follows 
\begin{align}\label{L1}
\mathcal{L}_{1}=\frac{1}{32\pi^{2}}\int_{0}^{\infty}\frac{ds}{s}e^{-\imath sm^{2}}\cot\left(\frac{sq_{(a)}z^{(a)}+sq_{(a)}\bar{z}^{(a)}}{2}\right)\cot\left(\frac{sq_{(a)}z^{(a)}-sq_{(a)}\bar{z}^{(a)}}{2}\right) \left[(q_{(a)}z^{(a)})^{2}-(q_{(a)}\bar{z}^{(a)})^{2}\right],
\end{align}
This can be easily verified by noting that $e^2(z^2-\bar{z}^2)=4\imath e^2 \mathcal{G}$, with $z=z^{(1)}$, and that the product of cotangents can be written in terms of $\Re \cos(sez)$ and $\Im \cos(sez)$, resulting in
\begin{equation}
     \mathcal{L}_{1}|_{g\to 0}=\frac{e^2}{32\pi^{2}}\int_{0}^{\infty}\frac{ds}{s}e^{-\imath sm^{2}}\frac{\Re \cos(sez)}{\Im \cos(sez)}F_{\mu\nu}\tilde{F}^{\mu\nu}.
\end{equation}

The Lagrangian \eqref{L1} contains the physical information of all one-loop diagrams with gauge bosons as external legs and is sufficient to obtain a range of phenomena. Nevertheless, it presents singularities in the small $s$ region, which can be traced back to the zero and two-point one-loop diagrams \cite{Schwartz:2014sze}, which we know to be UV-divergent from perturbation theory. To renormalize it, we notice that the expansion of the integrand around weak fields is singular up to second order in $s$
\begin{align}
    \mathcal{L}_{1}=\frac{1}{32\pi^{2}}\int_{0}^{\infty}\frac{ds}{s}e^{-\imath sm^{2}}\left\{\frac{4}{s^2}-\frac{2}{3}\left[(q_{(a)}z^{(a)})^{2}+(q_{(a)}\bar{z}^{(a)})^{2} \right] + \mathcal{O}(s^2) \right\},
\end{align}
and, therefore, these terms need to be regulated by the introduction of appropriate counterterms, which are given by the two divergent terms of the perturbative expansion. The weak field expansion of the renormalized Lagrangian is 
\begin{align}
\mathcal{L}_{1}\simeq\frac{1}{32\pi^{2}}\int_{0}^{\infty}\frac{ds}{s}e^{-\imath sm^{2}}\left\{-\frac{s^{2}}{180}\left[3(q_{(a)}z^{(a)})^{4}-22(q_{(a)}z^{(a)})^{2}(q_{(a)}\bar{z}^{(a)})^{2}+3(q_{(a)}\bar{z}^{(a)})^{4}\right]+\mathcal{O}(s^4)\right\}.
\end{align}

At last, evaluating the proper-time integral, which can be done via a Laplace transform, and rewriting the expression in terms of the Lorentz scalars $\mathcal{F}^{(a)}$ and $\mathcal{G}^{(a)}$, we get the effective Lagrangian for the weak field expansion
\begin{align}\label{correcaoquantica}
\mathcal{L}_{eff}&=\sum_{(a)=1}^{2}\left\{-\mathcal{F}^{(a)}+C_{1}^{(a)}(\mathcal{F}^{(a)})^{2}+C_{2}^{(a)}(\mathcal{G}^{(a)})^{2}\right.\nonumber\\
&+\frac{1}{2}K_{(a\bar{b})}\left[\mathcal{F}^{(a)}\mathcal{F}^{(\bar{b})}+10\mathcal{G}^{(a)}\mathcal{G}^{(\bar{b})}+11\Big((\mathcal{F}^{(a)})^{2}+(\mathcal{G}^{(a)})^{2}\Big)^{1/2}\Big((\mathcal{F}^{(\bar{b})})^{2}+(\mathcal{G}^{(\bar{b})})^{2}\Big)^{1/2}\right]\\
&+\left.N_{(a)}\left[11\Big(\mathcal{F}^{(a)}+\imath\mathcal{G}^{(a)}\Big)\Big(\mathcal{F}^{(a)}-\imath\mathcal{G}^{(a)}\Big)^{1/2}\Big(\mathcal{F}^{(\bar{b})}-\imath\mathcal{G}^{(\bar{b})}\Big)^{1/2} -3\Big(\mathcal{F}^{(a)}+\imath\mathcal{G}^{(a)}\Big)^{3/2}\Big(\mathcal{F}^{(\bar{b})}+\imath\mathcal{G}^{(\bar{b})}\Big)^{1/2} + c.c.\right]\right\},\nonumber
\end{align}
where we denote by $(\bar{b})$ the indices which are different from a given fixed index $(a)$ (so that $(\bar{b})\neq (a)$). The constants in \eqref{correcaoquantica} are defined according to
\begin{align}
C_{1}^{(a)}=\frac{8}{45}\frac{\alpha_{(a)}^{2}}{m^{4}},~~~~~C_{2}^{(a)}=\frac{14}{45}\frac{\alpha_{(a)}^{2}}{m^{4}},~~~~~K_{(a\bar{b})}=\frac{4}{45}\frac{\alpha_{(a)}\alpha_{(\bar{b})}}{m^{4}},~~~~~N_{(a)}=\frac{2\alpha_{(12)}\alpha_{(a)}}{45m^{4}},
\end{align}
with $\alpha_{(a)}=\frac{q^{2}_{(a)}}{4\pi}$ and $\alpha_{(12)}=\sqrt{\alpha_{(1)}\alpha_{(2)}}$.

Then, the Lagrangian in Eq. \eqref{correcaoquantica} contains the leading nontrivial contributions describing nonlinear electromagnetic self-interactions of light. In particular, the first line can be identified as an Euler-Heisenberg-type term for both photons and metaphotons. The remaining terms represent mixing photon-metaphoton interaction vertices, introducing additional complexity and leading to a richer structure compared to QED.

Moreover, the lagrangian \eqref{correcaoquantica} preserves the invariance of the classical action under P and T transformations, contrary to the approach of \cite{Kovalevich:1997de}. This is easily seen by noticing that potentially violating terms containing $\mathcal{G}$ all come as squares or complex conjugate pairs. Beyond the weak field approximation, expression \eqref{L1} is also invariant, which can be seen by the exchange $z\leftrightarrow\bar{z}$ and remembering that the magnetic charge $g$ is a pseudo-scalar. Therefore, the full one-loop effective action preserves the classical action discrete symmetries.

\subsection{Schwinger Pair Production}
Beyond the divergences coming from the perturbative expansion, the Lagrangian \eqref{L1} develops singularities for some specific values of the electromagnetic fields. To locate such singularities, we first consider the case of parallel fields $\vec{E}_{(a)}\parallel\vec{B}_{(a)}$. After some algebraic manipulations, the expression can be written as
\begin{equation}
    \mathcal{L}_1=-\frac{1}{8\pi^2}\int_0^\infty \frac{ds}{s}e^{-sm^2}(q_{(1)}E_{(1)}+q_{(2)}E_{(2)})(q_{(1)}B_{(1)}+q_{(2)}B_{(2)})\coth(sq_{(a)}B_{(a)})\cot(sq_{(a)}E_{(a)}),
\end{equation}
where we took $s\to -is$ to make the expression more clearly real. The expansion of $\coth(sq_{(a)}B_{(a)})$ yields a fully regular expression, showing that no poles occur for any finite value of $B_{(a)}$.
In particular, we can take the limit $B_{(a)}\to 0$, explicitly showing that all singularities are associated with the electric fields $E_{(a)}$. In this way, we see that
\begin{equation}
s_{n}=\frac{n\pi}{q_{(1)}E_{(1)}+q_{(2)}E_{(2)}},~~~~~~n=1,2,\ldots,
\end{equation}
corresponding to the poles of the cotangent function. These poles produce an imaginary part in the otherwise real effective Lagrangian, representing an instability of the vacuum under strong electric fields, which signals the onset of the Schwinger pair production mechanism \cite{Dunne:2025cyo}. This indicates that strong electric fields can induce the creation of dyon-antidyon pairs in dQED, in close analogy with the pair-production prediction of the Euler-Heisenberg action of QED.
The pair production rate is directly related to the imaginary part of the effective Lagrangian, $\Im(\mathcal{L}_{eff})$, which is given by the sum over the poles at $s=s_n$
\begin{equation}
\Im(\mathcal{L}_{eff})=\frac{\big(q_{(1)}E_{(1)}+q_{(2)}E_{(2)}\big)^{2}}{8\pi^{3}}\sum_{n=1}^{\infty}\frac{1}{n^{2}}\exp[-\frac{n\pi m^{2}}{(q_{(1)}E_{(1)}+q_{(2)}E_{(2)})}].
\end{equation}

For the regime $(q_{(1)}E_{(1)}+q_{(2)}E_{(2)})\ll m^{2}$, the rate of pair production is negligible, until it reaches an analog of the Schwinger limit for the dQED. In the case of QED, these fields are enormous and have not yet been produced in a laboratory, so that this phenomenon could be observed in a controlled environment. In our case, it seems reasonable to expect these fields also to be strong, given the relation above. Once again, at the limit where the magnetic charge vanishes, one recovers the usual Euler-Heisenberg result of QED. However, a distinctive feature of dQED emerges when the electric field in the $e$-sector vanishes: taking $E_{(1)}\to 0$, pair production still arises due to the contribution of the $g$-sector, $E_{(2)}$. Therefore, from \eqref{fieldsE_B}, one observes that in the dQED the magnetic sector effectively contributes to pair production\footnote{This was already known for the case of scalar monopoles studied in \cite{Affleck:1981ag}}.
\section{Dispersion Relations}
\label{IV}
In order to get the dispersion relations for the effective Lagrangian \eqref{correcaoquantica}, we first obtain its equations of motion. Applying the action principle to \eqref{correcaoquantica}, we get the field equations 
\begin{align}\label{eqdisplac}
		\partial_{\mu}\mathcal{D}^{\mu\nu}_{(c)}=0,
\end{align}
where $\mathcal{D}^{\mu\nu}_{(c)}$ is the electromagnetic displacement tensor (see Appendix \ref{apx:B} for the explicit derivation). This equation is accompanied by the usual Bianchi identity $\partial_{\mu}\tilde{F}_{(c)}^{\mu\nu}=0$ for each sector. The equation \eqref{eqdisplac} represents the vacuum field equations for the dyon Euler-Heisenberg electrodynamics. Its components are 
\begin{align}
		\mathcal{D}^{i0}_{(c)}=(\vec{D}_{(c)})^{i},~~~~~~~~~~~~\mathcal{D}^{ij}_{(c)}=-\epsilon^{ijk}(\vec{H}_{(c)})^{k}
\end{align}
with $\epsilon^{ijk}$ being the completely antisymmetric Levi-Civita symbol and $\vec{D}_{(c)}$ and $\vec{H}_{(c)}$ analogs of the electric displacement and magnetic field strength, respectively.

With this, we can write the corresponding equations
of motion as
\begin{align}
&\vec{\nabla}\cdot\vec{D}_{(c)}=0,~~~~~~\vec{\nabla}\times\vec{H}_{(c)}-\partial_{t}\vec{D}_{(c)}=0,\nonumber \\
&\vec{\nabla}\cdot\vec{B}_{(c)}=0,~~~~~~\vec{\nabla}\times\vec{E}_{(c)}+\partial_{t}\vec{B}_{(c)}=0.\label{nlmaxwell}
\end{align}
It is worth noting that we employ here the same decomposition of the gauge-sector components into electric and magnetic fields $(\vec{E}_{(a)},\vec{B}_{(a)})$ that was previously used to construct the Lorentz invariants.

Having established the Lagrangian and the corresponding field equations that describe the Euler-Heisenberg electrodynamics for dyons, we now turn to additional aspects of the theory, particularly the electromagnetic properties of the vacuum. The first step in this direction is to analyze the dispersion relations of electromagnetic waves propagating through this modified vacuum. In the following, we examine this aspect in more detail.

The Lagrangian \eqref{correcaoquantica} exhibits an involved interplay between the two gauge sectors, which makes it difficult to analyze the complete general case. Nevertheless, following the same reasoning adopted in \cite{Dittrich:1998gt,Kruglov:2007bh,Kruglov:2007zr} for QED, we begin by analyzing a simpler situation. In this approach, we decompose the electromagnetic fields $(\vec{E}{^{(a)}}, \vec{B}^{(a)})$ into strong, constant background fields $(\vec{E}_{0}^{(a)}, \vec{B}_{0}{^{(a)}})$ and small perturbations associated with a propagating wave $(\vec{e}^{(a)}, \vec{b}^{(a)})$. For simplicity, we restrict the analysis to a purely magnetic background, setting the electric components in both gauge sectors to zero, i.e., $\vec{E}^{(a)}_{0} = 0$ for all $(a)$. Therefore, the decomposition of the fields is given by
\begin{align}
\vec{E}_{(c)}=\vec{e}_{(c)},~~~~~~~~~~~~~~~~~\vec{B}_{(c)}=\vec{B}_{0(c)}+\vec{b}_{(c)}.
\end{align}

The constitutive equations for $\vec{D}_{(c)}$ and $\vec{H}_{(c)}$ are
\begin{align}
D^{k}_{(c)}=\frac{\partial\mathcal{L}_{eff}}{\partial E^{k}_{(c)}},~~~~~~~~~~H^{k}_{(c)}=-\frac{\partial\mathcal{L}_{eff}}{\partial B^{k}_{(c)}},
\end{align}
so that linearizing~\cite{Spallicci:2024drl} these equations with respect to the wave fields $\vec{e}_{(c)}$ and $\vec{b}_{(c)}$, we find the permittivity and permeability tensor of the dQED Euler-Heisenberg theory

\begin{align}\label{deh}
d^{(c)}_{i}=\varepsilon_{ij}e^{(c)}_{j}+\rho^{(c\bar{a})}_{ij}e^{(\bar{a})}_{j},~~~~~~~~~~~~h^{(c)}_{i}=(\mu^{-1})_{ij}b^{(c)}_{j}+\sigma^{(c\bar{a})}_{ij}b^{(\bar{a})}_{j},
\end{align}
with the diagonal and off-diagonal terms given by
\begin{align}\label{permtensors}
&\varepsilon_{ij}=\beta_{(c)}\delta_{ij}+\gamma_{(c)}B_{0i}^{(c)}B_{0j}^{(c)},~~~~~~~~~~~~~~~~~~~~~~~~~(\mu^{-1})_{ij}=\beta_{(c)}\delta_{ij}-\lambda_{(c)}B_{0i}^{(c)}B_{0j}^{(c)}\nonumber\\
&\\
&\rho^{(c\bar{a})}_{ij}=\rho_{0}^{(c)}B_{0i}^{(c)}B_{0j}^{(\bar{a})},~~~~~~~~~~~~~~~~~~~~~~~~~~~~~~~~~~~~~~\sigma^{(c\bar{a})}_{ij}=-\sigma_{0}^{(c)}B_{0i}^{(c)}B_{0j}^{(\bar{a})},\nonumber
\end{align}
and the coefficients defined as
\begin{align}\nonumber
&\beta_{(a)}=1-C_{1}^{(a)}B_{0(a)}^{2}-6K_{(a\bar{b})}B_{0(\bar{b})}^{2}-4N_{(\bar{b})}\frac{B_{0(\bar{b})}^{3}}{B_{0(a)}}-12N_{(a)}B_{0(a)}B_{0(\bar{b})}\\\nonumber
&\gamma_{(a)}=2C_{2}^{(a)}+11K_{(a\bar{b})}\frac{B_{0(\bar{b})}^{2}}{B_{0(a)}^{2}}+32N_{(a)}\frac{B_{0(\bar{b})}}{B_{0(a)}}+4N_{(\bar{b})}\frac{B_{0(\bar{b})}^{3}}{B_{0(a)}^{3}},\\\label{coef}
&\lambda_{(a)}=2C_{1}^{(a)}-4N_{(\bar{b})}\frac{B_{0(\bar{b})}^{3}}{B_{0(a)}^{3}}+12N_{(a)}\frac{B_{0(\bar{b})}}{B_{0(a)}},\\\nonumber
&\rho_{0}^{(a)}=10K_{(a\bar{b})}+10N_{(a)}\frac{B_{0(a)}}{B_{0(\bar{b})}}+N_{(\bar{b})}\frac{B_{0(\bar{b})}}{B_{0(a)}},\\\nonumber
&\sigma_{0}^{(a)}=12K_{(a\bar{b})}+12N_{(a)}\frac{B_{0(a)}}{B_{0(\bar{b})}}+21N_{(\bar{b})}\frac{B_{0(\bar{b})}}{B_{0(a)}}.
\end{align}
with no implicit summation on the gauge sector indices of the tensors \eqref{permtensors} and of the coefficients \eqref{coef}.

From equations \eqref{deh}-\eqref{coef}, we can see the effect of mixing terms between the two sectors, which will influence the propagation of the gauge fields $A_\mu$ and $B_\mu$. These effects disappear in the limit of vanishing magnetic charge, in which the magnetic sector decouples and we recover the results from QED. Conversely, if the electric charge is turned off, the theory describes a QED-like framework where the roles of the electric and magnetic charges are interchanged. This behavior is precisely what one would expect, since the model is symmetric under the exchange of the two charges~\cite{Dias:2025xwu,Govaerts:2023iqf}.

To explore the dispersion relations (DRs) associated with an electromagnetic wave propagating in the presence of an external background field, we decompose the electromagnetic fields $\vec{e}_{(a)}$ and $\vec{b}_{(a)}$ into plane waves~\cite{Gaete:2017cpc,Neves:2022jqq}:
\begin{align}
\vec{e}_{(a)}(\vec{r},t)=\vec{e}_{0(a)}e^{\imath(\vec{k}\cdot\vec{r}-\omega t)},~\text{and}~\vec{b}_{(a)}(\vec{r},t)=\vec{b}_{0(a)}e^{\imath(\vec{k}\cdot\vec{r}-\omega t)}
\end{align}
in which the electric $\vec{e}_{0(a)}$ and magnetic $\vec{b}_{0(a)}$ amplitudes are related by the Maxwell equations \eqref{nlmaxwell}, such that $\omega\vec{b}_{0(a)}=\vec{k}\times\vec{e}_{0(a)}$, with $\vec{k}$ denoting the wave vector and $\omega$ the frequency. Substituting these solutions in the field equations, the wave equation of the electric amplitude is
\begin{equation}\label{waveeq}
M_{ij}^{(ca)}e^{(a)}_{0j}=0,
\end{equation}
in which the matrix elements of $M_{ij}^{(ca)}$ are written as
\begin{align}\label{matrix}
M_{ij}^{(ca)}=&\left[(\omega^{2}-k^{2})\beta_{(a)}\delta_{ij}+\beta_{(a)}k_{i}k_{j}+\omega^{2}\gamma_{(a)}B_{0i}^{(a)}B_{oj}^{(a)}+\lambda_{(a)}(\vec{k}\times\vec{B}_{0}^{(a)})_{i}(\vec{k}\times\vec{B}_{0}^{(a)})_{j} \right ] \delta^{(ca)}+\\
&+\omega^{2}\rho_{0}^{(c)}B_{oi}^{(c)}B_{oj}^{(\bar{a})}+\sigma_{0}^{(c)}(\vec{k}\times\vec{B}_{0}^{(c)})_{i}(\vec{k}\times\vec{B}_{0}^{(\bar{a})})_{j},\nonumber
\end{align}
where in the first line we have only diagonal terms, i.e., only contributions from the same gauge sector\footnote{Once again we assert that there is no implicit summation on the gauge sector indices.}, while in the second line we have off-diagonal terms representing mixing terms. The dispersion relations are determined by the condition $\det(M)=0$. In this case, we have a block matrix with respect to gauge sectors $(a)=1,2$, so the determinant is obtained by
\begin{align}
\det(M)=\det(M^{(aa)})\det(M^{(\bar{b}\bar{b})}-M^{(\bar{b}a)}[M^{(aa)}]^{-1}M^{(a\bar{b})}).
\end{align}
Assuming that $M^{(aa)}$, with $(a)=1,2$, is invertible, the dispersion relation is then defined by the null determinant condition 
\begin{align}
\det(M^{\bar{(bb)}}-M^{(\bar{b}a)}[M^{(aa)}]^{-1}M^{(a\bar{b})})=0.
\end{align}
The inverse $[M^{(aa)}]^{-1}$ can be written as
\begin{equation}
[M^{(aa)}_{ij}]^{-1}=\eta_{(a)}\delta_{ij}+\chi_{(a)}k_{i}k_{j}+\theta_{(a)}B_{0i}^{(a)}B_{0j}^{(a)}+\tau_{(a)}(\vec{k}\times\vec{B}_{0}^{(a)})_{i}(\vec{k}\times\vec{B}_{0}^{(a)})_{j}+\xi_{(a)}(k_{i}B_{0j}^{(a)}+B_{0i}^{(a)}k_{j}),
\end{equation}
so that, using the fact that $M^{(aa)}_{im}[M^{(aa)}_{mj}]^{-1}=\delta_{ij}$, the inverse matrix coefficients are determined as 
\begin{align}
\eta_{(a)}=&\frac{1}{(\omega^{2}-k^{2})\beta_{(a)}},\\
\chi_{(a)}=&-\frac{(\omega^{2}-k^{2})\beta_{(a)}+\omega^{2}\gamma_{(a)}B_{0(a)}^{2}}{\omega^{2}(\omega^{2}-k^{2})\beta_{(a)}[(\omega^{2}-k^{2})\beta_{(a)}+(\omega^{2}B_{0(a)}^{2}-(\vec{k}\cdot\vec{B}_{0(a)})^{2})\gamma_{(a)}]},\\
\theta_{(a)}=&-\frac{\omega^{2}\gamma_{(a)}}{(\omega^{2}-k^{2})\beta_{(a)}[(\omega^{2}-k^{2})\beta_{(a)}+(\omega^{2}B_{0(a)}^{2}-(\vec{k}\cdot\vec{B}_{0(a)})^{2})\gamma_{(a)}]},\\
\tau_{(a)}=&-\frac{\lambda_{(a)}}{(\omega^{2}-k^{2})\beta_{(a)}[(\omega^{2}-k^{2})\beta_{(a)}+(k^{2}B_{0(a)}^{2}-(\vec{k}\cdot\vec{B}_{0(a)})^{2})\lambda_{(a)}]},\\
\xi_{(a)}=&\frac{(\vec{k}\cdot\vec{B}_{0(a)})\gamma_{(a)}}{(\omega^{2}-k^{2})\beta_{(a)}[(\omega^{2}-k^{2})\beta_{(a)}+(\omega^{2}B_{0(a)}^{2}-(\vec{k}\cdot\vec{B}_{0(a)})^{2})\gamma_{(a)}]}.
\end{align}
With this, the corresponding DRs are 
\begin{align}\label{generaldisprel}
    &\omega^2 \beta_{(a)} \left( (\omega^2 - k^2)\beta_{(a)} + \lambda_{(a)} \left(k^2 B_{0(a)}^2 - (\vec{k} \cdot \vec{B}_{0(a)})^2\right) - \frac{\rho^{(a)} \rho^{(\bar{b})} \left(\omega^2 B_{0(a)}^2 - (\vec{k} \cdot \vec{B}_{0(a)})^2\right)\left( \omega^2 B_{0(\bar{b})}^2 - (\vec{k} \cdot \vec{B}_{0(\bar{b})})^2\right)}{(\omega^2 - k^2)\beta_{(\bar{b})} + \gamma_{(\bar{b})} \left(\omega^2 B_{0(\bar{b})}^2 - (\vec{k} \cdot \vec{B}_{0(\bar{b})})^2\right) } \right)\nonumber\\
    &\times \left( (\omega^2 - k^2)\beta_{(a)}  + \gamma_{(a)} \left(\omega^2 B_{0(a)}^2 - (\vec{k} \cdot \vec{B}_{0(a)})^2\right) - \frac{\sigma^{(a)}  \sigma^{(\bar{b})} \left(k^2 B_{0(a)}^2 - (\vec{k} \cdot \vec{B}_{0(a)})^2\right)\left(k^2 B_{0^(\bar{b})}^2 - (\vec{k} \cdot \vec{B}_{0(\bar{b})})^2\right)}{(\omega^2 - k^2)\beta_{(\bar{b})} + \lambda_{(\bar{b})} \left(k^2 B_{0(\bar{b})}^2 - (\vec{k} \cdot \vec{B}_{0(\bar{b})})^2\right)} \right)  = 0.
\end{align}

Although obtaining an analytical solution for this expression in the form $\omega = \omega(k)$ is quite daunting, some special cases can be treated analytically and may provide some useful insights. When both charges vanish, $q_{(1)} = q_{(2)} = 0$, the DRs for both gauge sectors become those of free fields. In this situation, the photon and the metaphoton satisfy
\begin{equation}
\omega^{2} - k^{2} = 0,
\end{equation}
which is the standard propagation of a free field in the vacuum.

When one of the charges is set to zero, we recover the case of QED (or its magnetic analog). For illustration, let us take $q_{(1)} \neq 0$ and $q_{(2)} = 0$. In this scenario, the $g$-sector recovers the free field equation
\begin{equation} \label{'qed'g-sec}
\omega^{2}-k^{2}=0,
\end{equation}
whose solutions $\omega^{2} = k^{2}$ correspond to the metaphoton propagation modes in the vacuum. However, the $e$-sector reduces to a solvable polynomial equation of fourth degree
\begin{equation}
     \left.\left( (\omega^2 - k^2)\beta_{(1)} + \lambda_{(1)} \left(k^2 B_{0(1)}^2 - (\vec{k} \cdot \vec{B}_{0(1)})^2\right) \right)
    \left( (\omega^2 - k^2)\beta_{(1)}  + \gamma_{(1)} \left(\omega^2 B_{0(1)}^2 - (\vec{k} \cdot \vec{B}_{0(1)})^2\right) \right)\right|_{q_{(2)}=0}  = 0,
\end{equation}
with the coefficients $\beta_{(1)},  \lambda_{(1)}$ and $\gamma_{(1)}$ reduced to
\begin{equation}
     \left.\beta_{(1)}\right|_{q_{(2)}=0} = 1-C_1^{(1)}B_{0(1)}^2, \quad  \left.\lambda_{(1)}\right|_{q_{(2)}=0} = 2C^{(1)}_1, \quad  \left.\gamma_{(1)}\right|_{q_{(2)}=0} =2C_2^{(1)}=\frac{7}{2}C^{(1)}_1
\end{equation}
while the coefficients of the metaphoton sector are trivial.

The corresponding solutions are the propagation modes $\omega^{2} = k^{2}\Omega_\pm(\theta,B_{0(1)})$, where

\begin{align}\label{omega}
\Omega_\pm(\theta,B_{0(1)})=&\frac{1}{\Big(1+\frac{3}{2}C^{(1)}_{1}B_{0(1)}^{2}-\frac{5}{2}(C^{(1)}_{1}B_{0(1)}^{2})^{2}\Big)}\Bigg\{1-\frac{5}{4}C^{(1)}_{1}B_{0(1)}^{2}-\frac{13}{4}(C^{(1)}_{1}B_{0(1)}^{2})^{2}+\nonumber\\
&+\Big(\frac{11}{4}+\frac{3}{4}C^{(1)}_{1}B_{0(1)}^{2}\Big)C_1^{(1)}B_{0(1)}^{2}\cos^{2}\theta \pm \Big(\frac{3}{4}-\frac{17}{4}C^{(1)}_{1}B_{0(1)}^{2}\Big)C^{(1)}_{1}B_{0(1)}^{2}\sin^2\theta\Bigg\}.
\end{align}

In this scenario, we can highlight the following cases: 
(i) For $\theta = 0$, i.e., $\vec{k} \parallel \vec{B}_{0}^{(1)}$, we find that $\Omega_\pm(0,B_{0(1)})=1$, which implies that we have a free field solution $\omega^{2} = k^{2}$, indicating that photon propagation is not affected.
(ii) For $\theta = \pi/2$, i.e., $\vec{k} \perp \vec{B}_{0}^{(1)}$, two distinct propagation modes arise, which depend on both the orientation and the polarization of the photon with respect to the external magnetic field. 
Regarding $\Omega_{+}|_{\theta = \pi/2}=\frac{1 - 3C_1^{(1)}B^2_{0(1)}}{1 - C_1^{(1)}B^2_{0(1)}}$, we must deal with a pole located at $B_{\text{crit}}^{2} = 1/C_1^{(1)}$, and a zero at $B_{\text{zero}}^{2} = 1/3C_1^{(1)}$. Accordingly, three regions can be identified: $\Omega_{+}|_{\theta = \pi/2} > 0$ for $0 < B_{0(1)}^2 < B_{\text{zero}}^{2}$; $\Omega_{+}|_{\theta = \pi/2} < 0$ for $B_{\text{zero}}^{2} < B_{0(1)}^{2} < B_{\text{crit}}^{2}$; and $\Omega_{+}|_{\theta = \pi/2} > 0$ for $B_{0(1)}^{2} > B_{\text{crit}}^{2}$.
As for $\Omega_{-}|_{\theta = \pi/2}=\frac{1 - C_1^{(1)}B_{0(1)}^2 }{1 + \frac{5}{2} C_1^{(1)}B_{0(1)}^2}$, there is no physical pole, since the apparent singularity at $B_{\text{crit}}^{2} = 1/C_1^{(1)}$ is removed. In this case, we obtain $\Omega_{-}|_{\theta = \pi/2} > 0$ for $0 < B_{0(1)}^{2} < B_{\text{crit}}^{2}$, and $\Omega_{-}|_{\theta = \pi/2} < 0$ for $B_{0(1)}^{2} > B_{\text{crit}}^{2}$. In the limit where the field approaches the critical value $B_{\text{crit}}^{2}$, the function $\Omega_{-}|_{\theta = \pi/2}$ tends to zero, marking the transition between the two regimes.

It is worth emphasizing that the critical value $B_\text{crit}^2=1/C_1^{(1)}$ signals the limit of validity of the weak-field expansion, that is, for fields approaching $B_{\text{crit}}^{2}$, the truncation of the Euler-Heisenberg Lagrangian at lowest order is no longer valid. These critical fields are extremely large compared to the magnetic fields currently achieved in a laboratory, which are typically of order $O(1)$ T \cite{Ejlli:2020yhk}. Using the electron mass, as consistent with the QED limit being discussed, yields a critical field of order $B_{\text{crit}}\sim10^{11}$ T, comparable to those expected at the surface of magnetars \cite{Kaspi:2017fwg}. Using instead the lower mass bounds for spinor dyons (see, e.g., \cite{Kovalevich:1997de}), it increases to fields of order $B_{\text{crit}}\sim10^{24}$ T, vastly exceeding the fields expected in extreme astrophysical events. A consistent description of magnetic fields approaching these scales would require higher-order terms to be taken into consideration in the effective Lagrangian expansion. Therefore, the weak-field expansion employed throughout this work is well justified for comparisons with the current experiments on vacuum birefringence, which will be discussed in the next section.

From these results, the existence of two distinct propagation modes $(\omega/k)^{2}$ indicates that the vacuum behaves as a birefringent medium. Within the range of validity of the weak-field expansion, the dispersion relations remain real, corresponding to propagating modes. As the background field approaches the critical scale, the dispersion relations develop an evanescent regime, indicating the onset of dichroism. However, near this threshold higher-order terms in the effective action expansion are necessary for a proper quantitative description. The same analysis applies when $q_{(1)} = 0$ and $q_{(2)} \neq 0$ upon exchange of the labels $(1) \rightarrow (2)$.
\section{Vacuum Birefringence}
\label{V}
The vacuum birefringence of the theory is obtained from the difference between the refractive indices for the perpendicular wave polarizations in relation to the direction of the magnetic background field. Conveniently, we can rewrite equation \eqref{matrix} in terms of refractive index components $n_{i}=k_{i}/\omega$~\cite{Santos:2023ooc},
\begin{align}\label{matrixnewform}
M_{ij}^{(ca)}=&\left[(1-n^{2})\beta_{(a)}\delta_{ij}+\beta_{(a)}n_{i}n_{j}+\gamma_{(a)}B_{0i}^{(a)}B_{0j}^{(a)}+\lambda_{(a)}(\vec{n}\times\vec{B}_{0}^{(a)})_{i}(\vec{n}\times\vec{B}_{0}^{(a)})_{j}\right]\delta^{(ca)}+\nonumber\\
&+\rho_{0}^{(c)}B_{0i}^{(c)}B_{0j}^{(\bar{a})}+\sigma_{0}^{(c)}(\vec{n}\times\vec{B}_{0}^{(c)})_{i}(\vec{n}\times\vec{B}_{0}^{(\bar{a})})_{j},
\end{align}
with $n=\sqrt{n_{i}n_{i}}$. As discussed in the last section, the general case coming from the solutions of the dispersion relation in \eqref{generaldisprel} is quite a challenging problem to face, so we restrict ourselves to a simpler situation, which will still present some interesting results. We fix the external magnetic fields on $z$-direction, $\vec{B}_{0}^{(a)}=B_{0}^{(a)}\hat{z}$ and the wave vector directed along the $x$-direction, $\vec{k}=k\hat{x}$ and assume a linear polarization of the wave on the $z$-axis, $\vec{e}_{0(a)}=e_{03}^{(a)}\hat{z}$, that is, a situation in which the external magnetic fields are parallel to the amplitude of the wave. Then, following the same reasoning of the last sections, in which we obtained the dispersion relations for $\omega(k)$, solving the equation \eqref{waveeq} with $M^{(ca)}_{ij}$ in terms of the refractive index $n$ from \eqref{matrixnewform}. The parallel refractive index $n_{\parallel}$ is then given by

\begin{align}\label{nparallelgeral}
    n_{\parallel}^\pm = & \left( 1 - \frac{1}{2\beta_{(1)}\beta_{(2)} } \left\{ -\left(\beta_{(1)}\gamma_{(2)}B_{0(2)}^2 + \beta_{(2)}\gamma_{(1)}B_{0(1)}^2\right) \pm \left[\left(\beta_{(1)}\gamma_{(2)}B_{0(2)}^2 - \beta_{(2)}\gamma_{(1)}B_{0(1)}^2\right)^2 +\right.\right.\right.  \nonumber \\
    & \left.  \left.\left. ~~~~~~~~~~~~~~~~~~~~~~~~~~+4\beta_{(1)}\beta_{(2)} \rho^{(1)}_{0}\rho^{(2)}_{0}B_{0(1)}^2 B_{0(2)}^2\right]^{\frac12}\right\}  \right)^\frac{1}{2}.
\end{align}

The second case to analyze is the linear polarization of the wave on the $y$-axis, $\vec{e}_{0(a)}=e_{02}^{(a)}\hat{y}$, \textit{i.e.} a situation in which the external magnetic fields are perpendicular to the amplitude of the wave. Now, the perpendicular refractive index $n_{\perp}$ is given by

\begin{align}\label{nperpgeral}
    n_{\perp}^\pm= & \Bigg( \frac{1}{2\big[(\beta_{(1)}-\lambda_{(1)}B_{0(1)}^2)(\beta_{(2)}-\lambda_{(2)}B_{0(2)}^2) -\sigma_{0}^{(1)}\sigma_{0}^{(2)}B_{0(1)}^2 B_{0(2)}^2\big]} \Big\{\beta_{(1)}(\beta_{(2)}-\lambda_{(2)}B_{0(2)}^2)+\beta_{(2)}(\beta_{(1)}-\lambda_{(1)}B_{0(1)}^2)\pm \nonumber \\
    & ~~~~~~ \pm \Big[ \Big(\beta_{(1)}(\beta_{(2)}-\lambda_{(2)}B_{0(2)}^2)-\beta_{(2)}(\beta_{(1)}-\lambda_{(1)}B_{0(1)}^2)\Big)^{2}+4\beta_{(1)}\beta_{(2)}\sigma^{(1)}_{0}\sigma^{(2)}_{0} B_{0(1)}^2 B_{0(2)}^2 \Big]^\frac{1}{2} \Big\} \Bigg)^\frac{1}{2}
\end{align}

Although a complete investigation of the refractive indices \eqref{nparallelgeral} and \eqref{nperpgeral} is a challenging task, several qualitative features can be extracted from their expressions. First, we can no longer identify distinct photon and metaphoton propagating modes, but only hybrid eigenmodes which are the true eigenvectors of the propagation matrix \eqref{matrixnewform}, due to the fermionic loop effectively coupling one gauge sector to the other. In this sense, it presents similarities with models for $U(1)$ dark sectors~\cite{deliyergiyev2016recent,bento2024classes,bauer2018hunting,Fayet:2024ddk,GomezSanchez:2011orv}. However, here the sectors are mixed via the one-loop effective theory rather than through a fundamental kinetic-mixing operator. Moreover, the parallel and perpendicular polarization modes are affected differently by the mixing terms, $n_{\parallel}^\pm$ with $\rho_0$ and $n_{\perp}^\pm$ with $\sigma_0$, due to the structure of \eqref{matrixnewform}. The two branches of the refractive indices ($+$ and $-$) correspond to the eigenvalues of the propagation matrix \eqref{matrixnewform} and reduce to photon and metaphoton modes in the limit of vanishing mixing between gauge sectors. In the following, we analyze a few limit cases which allow for a more concrete evaluation of the results.

A useful simplification arises in a regime where the mixing between photons and metaphotons is strongly suppressed. This occurs, for instance, when there is a strong hierarchy between the charges, $q_{(1)}\gg q_{(2)}$ (or vice versa), so that one sector provides the dominant contribution to the effective action. Since the mixing coefficients are proportional to products of the two charges, they become parametrically suppressed whenever one charge is much smaller than the other and may therefore be neglected as a controlled approximation. Such a hierarchy can naturally arise in scenarios involving monopoles obeying the Dirac quantization condition\footnote{We remind the reader that this is not an assumption of the present model and that this example is introduced solely for illustrative purposes.} or millicharged particles associated with dark sectors. This regime should, however, be distinguished from exact decoupling, which is only achieved when one of the charges vanishes. In that case, the fermionic loop does not induce mixing, whereas for nonzero charges it generically couples both sectors. The charge hierarchy therefore defines a nearly decoupled regime rather than a strictly decoupled one.

This approximation can be implemented by taking $\rho_{0}^{(a)}\approx 0$ and $\sigma_{0}^{(a)}\approx 0$. As follows from \eqref{deh}, these parameters control the mixing between the sectors, and their suppression ensures that propagation is effectively governed by a single dominant sector. Under this assumption, the refractive indices reduce to the simpler form

\begin{equation}\label{nmixsup}
n_{\parallel}^{(a)}=\sqrt{1+\frac{\gamma_{(a)}}{\beta_{(a)}}B_{0(a)}^{2}},~~~~~~~~~~~~~n_{\perp}^{(a)}=\sqrt{\frac{\beta_{(a)}}{\beta_{(a)}-\lambda_{(a)}B_{0(a)}^{2}}}.
\end{equation}

Comparing the mixing-suppressed refractive indices \eqref{nmixsup} with the general expressions \eqref{nparallelgeral} and \eqref{nperpgeral}, we find that the mixing parameters $\rho_0^{(a)}$ and $\sigma_0^{(a)}$ are responsible for lifting the degeneracy between the propagating modes. In their absence, the hybrid propagation modes reduce to independent photon and metaphoton modes. To illustrate this mechanism, we consider the symmetric case $q_{(1)}=q_{(2)}$, although the same behavior holds in the general case. By varying the external field $B_{0(1)}$ while keeping $B_{0(2)}$ fixed, the decoupled refractive indices \eqref{nmixsup} become degenerate in the limit $B_{0(1)}\to B_{0(2)}$, while \eqref{nparallelgeral},\eqref{nperpgeral} reduce to
\begin{equation}\label{symmetric}
    n_{\parallel}^\pm =  \sqrt{ 1 - \frac{(\gamma_{(1)} \pm  \rho_{0}^{(1)})B_{0(1)}^2}{\beta_{(1)}} }, \quad\quad     n_{\perp}^\pm=  \sqrt{\frac{\beta_{(1)}[(\beta_{(1)}-\lambda_{(1)} B_{0(1)}^2)\pm 
      \sigma_{0}^{(1)} B_{0(1)}^2]}{\big[(\beta_{(1)}-\lambda_{(1)} B_{0(1)}^2)^2 -(\sigma_{0}^{(1)}B_{0(1)}^{2})^2 \big]}}.
\end{equation}

These results indicate that the vacuum birefringence $\delta n^\pm=n_{\parallel}^\pm - n_{\perp}^\pm $, in the weak field expansion, differs for each mode as
\begin{align}
&\delta n^{+} \simeq -\frac{176\alpha_{(1)}^{2}}{45m^{4}}B_{0(1)}^{2}-\frac{24562\alpha_{(1)}^{4}}{2025m^{8}}B_{0(1)}^{4}+O(B_{0(1)}^{6})\nonumber\\
&\delta n^{-} \simeq -\frac{3362\alpha_{(1)}^{4}}{2025m^{8}}B_{0(1)}^{4}+O(B_{0(1)}^{6}).
\end{align}
We observe a nontrivial hierarchy between the birefringent response of the two propagation modes. For the negative mode ($\delta n^{-}$), contributions to the birefringence cancel exactly at quadratic order in the external field, so that the first non-vanishing correction arises at quartic order, $O(B_{0(1)}^{4})$. In contrast, the positive mode ($\delta n^{+}$) exhibits the standard behavior expected from an Euler-Heisenberg theory, with leading birefringent corrections appearing at quadratic order in the external field. This hierarchy indicates that the two propagation modes respond differently to the nonlinear structure of the effective action: while the positive mode exhibits the quadratic behavior characteristic of Euler-Heisenberg electrodynamics, the birefringence of the negative mode is protected against the leading quadratic-order corrections and becomes sensitive only to higher-order operators. Such a distinction disappears in the decoupled limit \eqref{nmixsup}, where the modes become degenerate. Consequently, the observation of two distinct birefringent responses could provide evidence for the existence of two distinct propagation modes ($n^\pm$), as predicted by the present model.

We now consider the QED limit discussed in the previous section. For ($q_{(1)} \neq 0$) and ($q_{(2)} = 0$), the mixing vanishes exactly and the coefficients \eqref{coef} reduce to 
\begin{align}
&\beta_{(1)}=1-C_{1}^{(1)}B_{0(1)}^{2},~~~~~\beta_{(2)}=1;\nonumber\\
&\gamma_{(1)}=2C_{2}^{(1)},~~~~~~~~~~~~~~~\gamma_{(2)}=0;\nonumber\\
&\lambda_{(1)}=2C_{1}^{(1)},~~~~~~~~~~~~~~~\lambda_{(2)}=0.
\end{align}
The refractive indices are then given by \eqref{nmixsup} for the photon sector, while the metaphoton sector becomes trivial, ($n^{(2)}_{\parallel}=n^{(2)}_{\perp}=1$). In this limit, the photon modes reduce to $n_{\parallel}^{(1)}=1/\sqrt{\Omega_{-}\vert_{\theta=\pi/2}}$ and $n_{\perp}^{(1)}=1/\sqrt{\Omega_{+}\vert_{\theta=\pi/2}}$. Considering the weak field expansion, motivated by the polarization vacuum with laser (PVLAS) experiment~\cite{Ejlli:2020yhk}, we get
\begin{equation} \label{birefinQED}
\delta n^{(1)}\simeq \frac{6\alpha_{(1)}^{2}}{45m^{4}}B_{0(1)}^{2}-\frac{146 \alpha^4_{(1)}}{2025m^8}B^4_{0(1)}+O(B_{0(1)}^6),
\end{equation}
which matches the result of QED at leading order $\Delta n_{\text{QED}}=\frac{6\alpha^{2}}{45m^{4}}B^{2}$~\cite{Ejlli:2020yhk, Sorokin:2021tge}. While the quartic correction suggests a slower growth of the birefringence as the external magnetic field increases, higher-order terms in the Euler-Heisenberg expansion must likewise be included once $O(B_{0(1)}^4)$ corrections become relevant, since they may also contribute to the birefringence. Therefore, in the special cases ($q_{(1)} \neq 0, q_{(2)} = 0$) and ($q_{(1)} = 0, q_{(2)} \neq 0$), the two sectors are effectively decoupled. The non-vanishing sector exhibits QED-like vacuum birefringence in the presence of the external magnetic field, while the other propagates trivially in vacuum. The relative propagation speed of the modes is therefore determined solely by which sector couples to the background field.
\section{Conclusions}
\label{VI}
Starting from a dyon quantum electrodynamics (dQED) model \cite{Govaerts:2023iqf,Dias:2025xwu}, inspired by the works of Cabibbo-Ferrari \cite{Cabibbo:1962td} and Salam \cite{Salam:1966bd}, and using Schwinger's proper-time method \cite{Schwinger:1951nm}, we derive an effective nonlinear electrodynamics (NLED) described by an Euler-Heisenberg-type Lagrangian \eqref{correcaoquantica}. This effective theory incorporates nonlinear contributions from both gauge sectors and reduces to the standard QED Euler-Heisenberg Lagrangian in the limit $g\to 0$. Conversely, in the limit $e\to 0$, the resulting theory also reproduces a QED-like Euler-Heisenberg Lagrangian for the magnetic sector, explicitly revealing the symmetry between the two gauge sectors. 

We analyze the analog of the Schwinger pair-production limit for the dQED, identifying values of the electric fields for which the effective Lagrangian develops a non-vanishing imaginary part. We then derive the nonlinear Maxwell equations for each gauge sector and the electromagnetic displacement tensor. The physical electric and magnetic fields \eqref{fieldsE_B} receive contributions from both gauge sectors, such that the resulting nonlinear theory exhibits a mixing between them.

To investigate the properties of the vacuum, we analyze the propagation of a plane electromagnetic wave in the presence of background magnetic fields. By linearizing the components of the displacement tensor, we determine the effective permittivity and permeability. These quantities allow us to derive the corresponding dispersion relations for the propagating modes of the theory, resulting in the general equation \eqref{generaldisprel}. The analysis of two limiting cases, when both charges vanish and when only one of the charges is non-vanishing, shows that the former reproduces the standard propagation of free fields in vacuum, while the latter recovers the Euler-Heisenberg result of QED for the non-trivial sector. This result indicates that the effective vacuum acts as a birefringent medium.

To explore the birefringence phenomenon, we derive the general refractive indices (\ref{nparallelgeral}) and (\ref{nperpgeral}) for the case where the background fields are perpendicular to the direction of wave propagation. In this case, the propagation modes are no longer identified with photons and metaphotons, but instead hybrid superpositions of both gauge sectors. We show that the parallel and perpendicular refractive indices involve different mixing parameters, $\rho_0$ and $\sigma_0$, respectively, thereby lifting the degeneracy of the propagating modes, as illustrated explicitly for the symmetric theory with $q_{(1)}=q_{(2)}$. We also identify a regime in which the mixing is parametrically suppressed, owing to the dominance of one sector over the other ($q_{(1)}\gg q_{(2)}$ or vice versa), leading to independent modes associated with each gauge sector. Finally, setting the magnetic charge to zero, the birefringence effects in the $e$-sector reduce to the well-known QED result \eqref{birefinQED}. Furthermore, we show that the refractive indices $n_{\perp}$ and $n_{\parallel}$ are modified by the presence of a magnetic charge, so that an experimental deviation from the QED prediction could signal the existence of a hidden magnetic charge~\cite{Fedotov:2022ely}.

The present framework provides a natural starting point for studying various nonlinear effects in the vacuum polarization structure of the theory, including a more complete analysis of the photon-metaphoton mixing modes and of vacuum dichroism. Other natural extensions include the photon-photon, photon-metaphoton and metaphoton-metaphoton scatterings, as well as the interaction energy between static sources within the model. On the gravitational side, coupling dQED to gravity may provide a natural framework for studying dyonic black-hole solutions and exploring possible astrophysical signatures~\cite{Abishev:2019nrb,Brihaye:1998cm}.
\subsection*{Acknowledgments}
The authors would like to thank Prof. J.A. Helay\"el-Neto for the discussions and guidance during the execution of the present work. D.O.R.A. acknowledges the PEDECIBA program and the ANII-FCE-2025-186497 project for financial support. CAPES-Brazil is acknowledged for invaluable financial help. 

\appendix

\section{\label{apx:B}Equations of Motion}
The action principle applied to the effective Lagrangian \eqref{correcaoquantica} yields~\cite{Dittrich:2000zu},
\begin{align}
0&=\frac{\partial\mathcal{L}_{eff}}{\partial A_{\nu}^{(a)}}-\partial_{\mu}\frac{\partial\mathcal{L}_{eff}}{\partial( \partial_{\mu}A_{\nu}^{(a)})}=\partial_{\mu}\frac{\partial\mathcal{L}_{eff}}{\partial( \partial_{\mu}A_{\nu}^{(a)})}=2\partial_{\mu}\frac{\partial\mathcal{L}_{eff}}{\partial F^{(a)}_{\mu\nu}}~~~~~~\Rightarrow~~~~~~2\partial_{\mu}\frac{\partial\mathcal{L}_{eff}}{\partial F^{(a)}_{\mu\nu}}=0,
\end{align}
and with the aid of the result
\begin{equation}
\frac{\partial\mathcal{F}^{(b)}}{\partial F^{(a)}_{\mu\nu}}=\frac{1}{2}F^{\mu\nu}_{(b)}\delta_{(ab)},~~~~~~~~~~\frac{\partial\mathcal{G}^{(b)}}{\partial F^{(a)}_{\mu\nu}}=-\frac{1}{2}\tilde{F}^{\mu\nu}_{(b)}\delta_{(ab)},
\end{equation}
where $\delta_{(ab)}$ is the Kronecker delta, the equations of motion are written as
\begin{align}
0&=2\partial_{\mu}\frac{\partial\mathcal{L}_{eff}}{\partial F^{(c)}_{\mu\nu}}=2\partial_{\mu}\left[\frac{\partial\mathcal{L}_{eff}}{\partial \mathcal{F}^{(b)}}\frac{\partial\mathcal{F}^{(b)}}{\partial F^{(c)}_{\mu\nu}}+\frac{\partial\mathcal{L}_{eff}}{\partial \mathcal{G}^{(b)}}\frac{\partial\mathcal{G}^{(b)}}{\partial F^{(c)}_{\mu\nu}}\right]\nonumber\\
&=\partial_{\mu}\left[\left(\frac{\partial\mathcal{L}_{eff}}{\partial \mathcal{F}^{(c)}}\right)F^{\mu\nu}_{(c)}-\left(\frac{\partial\mathcal{L}_{eff}}{\partial \mathcal{G}^{(c)}}\right)\tilde{F}^{\mu\nu}_{(c)}\right].
\end{align}
Taking the derivative in effective Lagrangian, we have the explicit result
\begin{align}
&0=\partial_{\mu}\Bigg\{\Big(1-2C_{1}^{(c)}\mathcal{F}^{(c)}\Big)F^{\mu\nu}_{(c)}-K_{(\bar{a}c)}\left[\mathcal{F}^{(\bar{a})}+11\Big(\big(\mathcal{F}^{(\bar{a})}\big)^{2}+\big(\mathcal{G}^{(\bar{a})}\big)^{2}\Big)^{1/2}\Big(\big(\mathcal{F}^{(c)}\big)^{2}+\big(\mathcal{G}^{(c)}\big)^{2}\Big)^{-1/2}\mathcal{F}^{(c)}\right]F^{\mu\nu}_{(c)}\nonumber\\
&+\frac{9}{2}N_{(c)}\left[\Big(\mathcal{F}^{(c)}+\imath\mathcal{G}^{(c)}\Big)^{1/2}\Big(\mathcal{F}^{(\bar{a})}+\imath\mathcal{G}^{(\bar{a})}\Big)^{1/2}+c.c.\right]F^{\mu\nu}_{(c)}\nonumber\\
&+\frac{3}{2}N_{(\bar{a})}\left[\Big(\mathcal{F}^{(\bar{a})}+\imath\mathcal{G}^{(\bar{a})}\Big)^{3/2}\Big(\mathcal{F}^{(c)}+\imath\mathcal{G}^{(c)}\Big)^{-1/2}+c.c.\right]F_{(c)}^{\mu\nu}\nonumber\\
&-11 N_{(c)}\left[\Big(\mathcal{F}^{(c)}-\imath\mathcal{G}^{(c)}\Big)^{1/2}\Big(\mathcal{F}^{(\bar{a})}-\imath\mathcal{G}^{(\bar{a})}\Big)^{1/2}+c.c.\right]F_{(c)}^{\mu\nu}\nonumber\\
&-\frac{11}{2}N_{(c)}\Bigg[\Big(\mathcal{F}^{(c)}+\imath\mathcal{G}^{(c)}\Big)\Big(\mathcal{F}^{(c)}-\imath\mathcal{G}^{(c)}\Big)^{-1/2}\Big(\mathcal{F}^{(\bar{a})}-\imath\mathcal{G}^{(\bar{a})}\Big)^{1/2}+c.c.\Bigg]F_{(c)}^{\mu\nu}\nonumber\\
&-\frac{11}{2}N_{(\bar{a})}\Bigg[\Big(\mathcal{F}^{(\bar{a})}+\imath\mathcal{G}^{(\bar{a})}\Big)\Big(\mathcal{F}^{(\bar{a})}-\imath\mathcal{G}^{(\bar{a})}\Big)^{1/2}\Big(\mathcal{F}^{(c)}-\imath\mathcal{G}^{(c)}\Big)^{-1/2}+c.c.\Bigg]F_{(c)}^{\mu\nu}\\
&+2C_{2}^{(c)}\mathcal{G}^{(c)}\tilde{F}^{\mu\nu}_{(c)}+K_{(\bar{a}c)}\left[10\mathcal{G}^{(\bar{a})}+11\Big(\big(\mathcal{F}^{(\bar{a})}\big)^{2}+\big(\mathcal{G}^{(\bar{a})}\big)^{2}\Big)^{1/2}\Big(\big(\mathcal{F}^{(c)}\big)^{2}+\big(\mathcal{G}^{(c)}\big)^{2}\Big)^{-1/2}\mathcal{G}^{(c)}\right]\tilde{F}^{\mu\nu}_{(c)}\nonumber\\
&-\frac{9}{2}N_{(c)}\left[\Big(\mathcal{F}^{(c)}+\imath\mathcal{G}^{(c)}\Big)^{1/2}\Big(\mathcal{F}^{(\bar{a})}+\imath\mathcal{G}^{(\bar{a})}\Big)^{1/2}+c.c.\right]\imath\tilde{F}^{\mu\nu}_{(c)}\nonumber\\
&-\frac{3}{2}N_{(\bar{a})}\left[\Big(\mathcal{F}^{(\bar{a})}+\imath\mathcal{G}^{(\bar{a})}\Big)^{3/2}\Big(\mathcal{F}^{(c)}+\imath\mathcal{G}^{(c)}\Big)^{-1/2}+c.c.\right]\imath\tilde{F}^{\mu\nu}_{(c)}\nonumber\\
&+11N_{(c)}\left[\Big(\mathcal{F}^{(c)}-\imath\mathcal{G}^{(c)}\Big)^{1/2}\Big(\mathcal{F}^{(\bar{a})}-\imath\mathcal{G}^{(\bar{a})}\Big)^{1/2}-c.c.\right]\imath\tilde{F}^{\mu\nu}_{(c)}\nonumber\\
&-\frac{11}{2}N_{(c)}\Bigg[\Big(\mathcal{F}^{(c)}+\imath\mathcal{G}^{(c)}\Big)\Big(\mathcal{F}^{(c)}-\imath\mathcal{G}^{(c)}\Big)^{-1/2}\Big(\mathcal{F}^{(\bar{a})}-\imath\mathcal{G}^{(\bar{a})}\Big)^{1/2}-c.c.\Bigg]\imath\tilde{F}^{\mu\nu}_{(c)}\nonumber\\
&-\frac{11}{2}N_{(\bar{a})}\Bigg[\Big(\mathcal{F}^{(\bar{a})}+\imath\mathcal{G}^{(\bar{a})}\Big)\Big(\mathcal{F}^{(\bar{a})}-\imath\mathcal{G}^{(\bar{a})}\Big)^{1/2}\Big(\mathcal{F}^{(c)}-\imath\mathcal{G}^{(c)}\Big)^{-1/2}-c.c.\Bigg]\imath\tilde{F}^{\mu\nu}_{(c)}\Bigg\}\nonumber.
\end{align}
The derivative term within the curly brackets can be recognized as the electromagnetic displacement tensor $\mathcal{D}^{\mu\nu}_{(c)}$. In this way, it can be expressed as a simple equation
\begin{equation}
    \partial_{\mu}\mathcal{D}^{\mu\nu}_{(c)}=0.
\end{equation}


\bibliography{refs}
\end{document}